\documentclass[conference,10pt,letterpaper]{IEEEtran}
\usepackage{cite}
\usepackage{amsmath,amssymb,amsfonts,graphicx}
\usepackage{booktabs}
\usepackage{bm}
\usepackage[table]{xcolor}
\usepackage{placeins}
\usepackage{fontawesome5}
\definecolor{arxivpink}{HTML}{D63384}
\usepackage[colorlinks=true,linkcolor=arxivpink,citecolor=arxivpink,urlcolor=arxivpink,filecolor=arxivpink,breaklinks=true]{hyperref}
\hypersetup{pdftitle={OSD-JSCC: Neural Joint Source--Channel Coding with One-Step Diffusion},pdfauthor={Shunpu Tang and Qianqian Yang}}

\title{\texorpdfstring{\makebox[0pt][r]{\raisebox{-0.2\height}[0pt][0pt]{\includegraphics[height=1.2em]{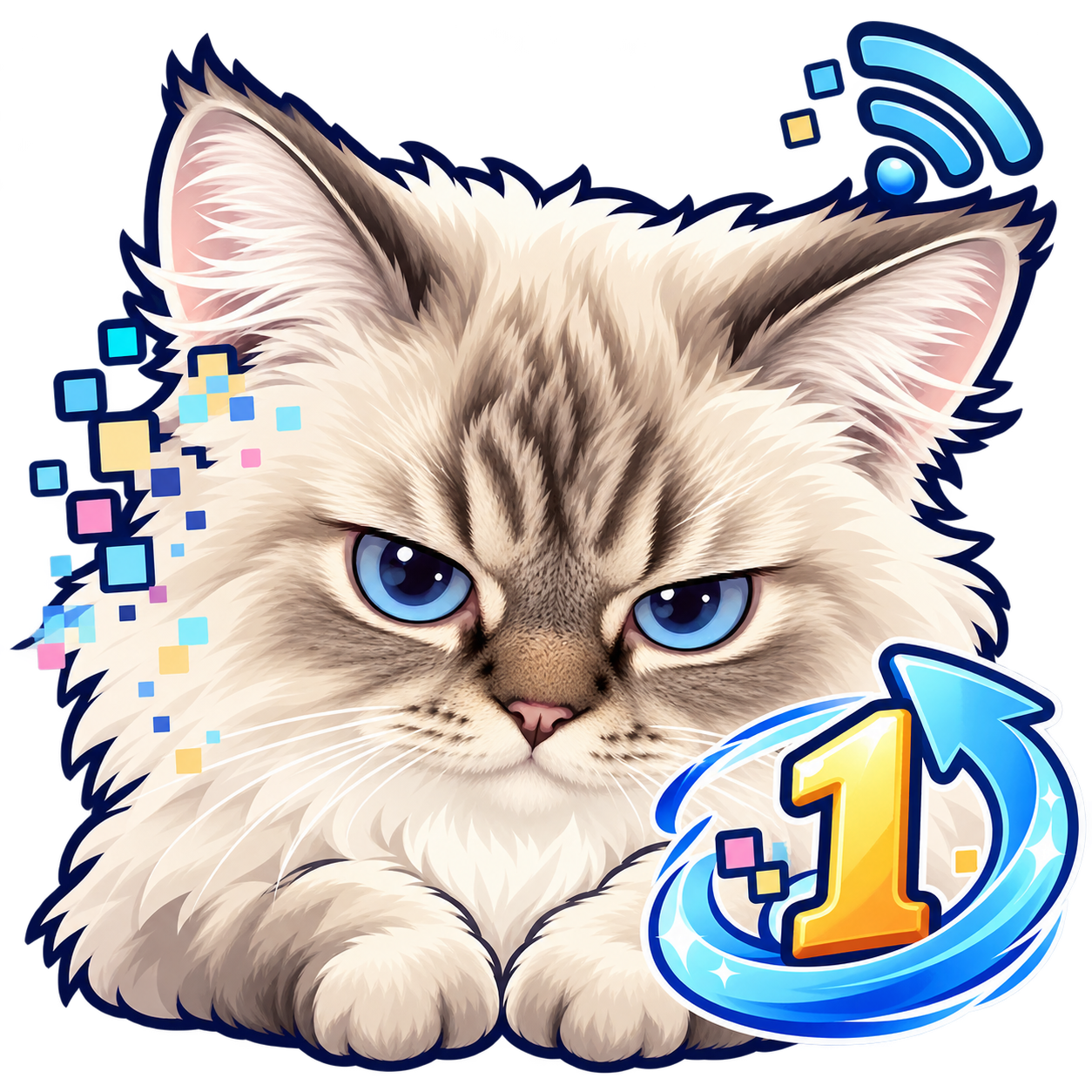}}\hspace{0.3em}}}{}OSD-JSCC: Neural Joint Source--Channel Coding with \texorpdfstring{\textcolor{red}{\textbf{O}}}{O}ne-\texorpdfstring{\textcolor{red}{\textbf{S}}}{S}tep \texorpdfstring{\textcolor{red}{\textbf{D}}}{D}iffusion}

\author{\IEEEauthorblockN{Shunpu Tang, Qianqian Yang}
\IEEEauthorblockA{College of Information Science and Electronic Engineering,
Zhejiang University, Hangzhou, China\\
\texttt{\{tanshunpu,qianqianyang20\}@zju.edu.cn}\\[2pt]
Code is available at \href{https://github.com/tangshunpu/OSD-JSCC}{\faGithub\hspace{0.35em}\texttt{github.com/tangshunpu/OSD-JSCC}}}}

\begin{document}

\maketitle

\begin{abstract}
Neural joint source–channel coding (JSCC) enables efficient image transmission over noisy channels, yet distortion-optimized approaches often produce blurry reconstructions. While diffusion priors can enhance visual fidelity, their iterative sampling incurs substantial decoding latency. To address this limitation, we propose OSD-JSCC, an SNR-adaptive JSCC framework with a single diffusion step. Specifically, OSD-JSCC transmits pretrained variational autoencoder (VAE) latents through an SNR-conditioned JSCC codec and incorporates two key designs. First, we introduce a stochastic diffusion-state interface that enables the JSCC decoder to predict a Gaussian diffusion state rather than a clean latent, while extracting image-specific semantic tokens. The sampled diffusion state preserves spatial structure and provides a good initialization for one-step refinement, whereas the semantic tokens, together with an SNR token, guide the reconstruction process. Second, we develop SNR-conditioned feature modulation and adaptive timestep selection to jointly adapt latent transmission and diffusion refinement to varying channel conditions. Furthermore, we adopt a three-stage training strategy with low-rank adaptation (LoRA) to efficiently adapt the pretrained diffusion prior to the JSCC task. Extensive experiments on Kodak and CLIC2020 demonstrate that OSD-JSCC achieves superior perceptual quality and decoding efficiency. Compared with existing diffusion-based JSCC baselines, OSD-JSCC reduces LPIPS by approximately 43\% on Kodak and 53\% on CLIC2020, while achieving up to a 50.6× reduction in decoding latency.

\end{abstract}


\section{Introduction}

The transmission of high-resolution images over noisy channels traditionally relies on a separate source and channel coding pipeline. Specifically, images are first compressed into bitstreams using conventional codecs, such as H.265/BPG and H.266/VVC, or learned image codecs~\cite{ballé2018variational,he2022elic}. The bitstreams are then protected by channel codes, such as LDPC or polar codes, before transmission. This design is supported by Shannon's separation theorem~\cite{shannon1948mathematical}, which allows source and channel coding to be optimized independently. However, its optimality requires sufficiently long blocklengths, which may be difficult to achieve under practical latency constraints. Moreover, under time-varying channel conditions, the selected source and channel coding rates may no longer match the instantaneous channel condition. Stronger channel coding increases redundancy, while insufficient protection may leave bit errors that corrupt the compressed bitstream and cause reconstruction failure.

To overcome these limitations, joint source--channel coding (JSCC) combines source and channel coding into a single end-to-end system. The encoder directly maps an image to channel symbols, while the decoder reconstructs the image from noisy channel observations. Recent neural JSCC methods~\cite{bourtsoulatze2019deep,choi2019neural,song2020infomax} use deep neural networks to learn these transformations and have shown promising performance under poor channel conditions. However, most methods focus on minimizing reconstruction distortion, often producing blurry images at high compression ratios and low signal-to-noise ratios (SNRs), which limits their perceptual quality.

\begin{figure}[!t]
\centering
\includegraphics[width=\columnwidth]{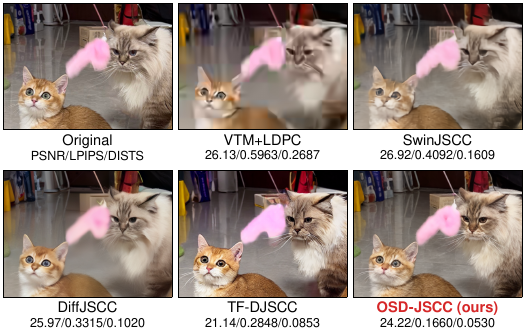}
\caption{Visual comparison of the proposed OSD-JSCC and the baseline methods on an AWGN channel at 0~dB.}
\label{fig:kodim23_awgn_0db}
\end{figure}

Recently, generative models have demonstrated remarkable capabilities in generating high-quality images and show great potential for optimizing rate-distortion-perception trade-offs \cite{blau2019rethinking} in image compression and reconstruction. In \textcolor{black}{the} JSCC domain, \textcolor{black}{prior works}~\cite{GAN_JSCC,shu} modeled the image transmission problem as \textcolor{black}{a} degradation process and proposed to use a generative adversarial network (GAN) to solve the inverse problem of reconstructing the original image. Recent works \cite{yang2024diffusion,letafati2025conditional,shunpu2025enabling} have explored the use of diffusion models for JSCC, showing promising results in terms of both reconstruction quality and perceptual fidelity. However, these diffusion-based methods typically require multiple iterative denoising steps during the decoding process, which leads to high computational costs and increased latency.

\begin{figure*}[!ht]
    \centering
    \includegraphics[width=\textwidth]{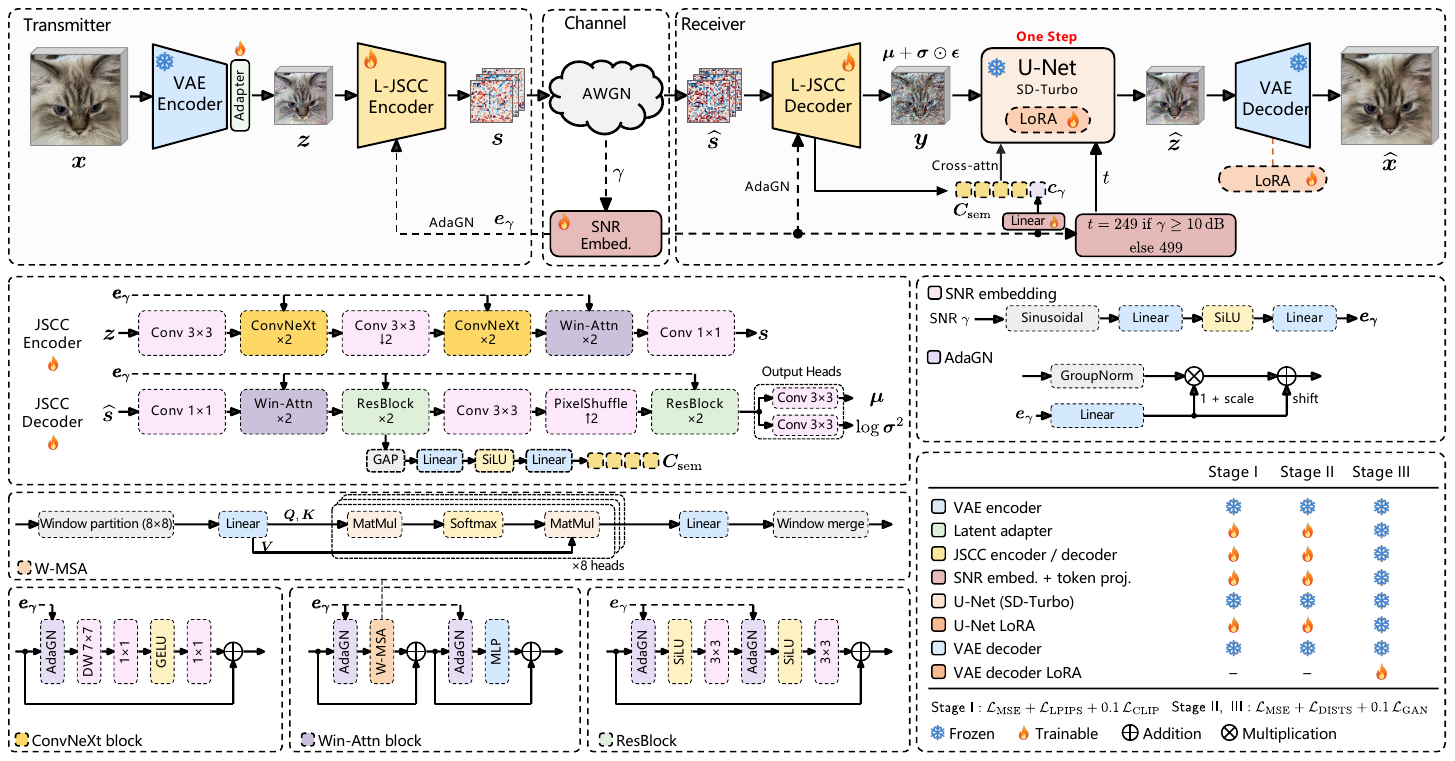}
    \caption{Overview of OSD-JSCC, including the transceiver, SNR-conditioned codec, constituent blocks, and three-stage training configuration.}
    \label{fig:osd_jscc_architecture}
\end{figure*}

Motivated by the recent success of one-step diffusion models, we explore their integration into the JSCC paradigm to achieve superior perceptual quality with substantially reduced decoding latency. Unlike one-step diffusion-based image compression, which reconstructs images from reliably decoded representations, JSCC must operate directly on stochastic and potentially severely corrupted channel observations. The key challenge is therefore to handle a broad range of channel noise levels in terms of signal-to-noise ratio (SNR), with the limited denoising capacity of a single diffusion step, while recovering high-quality and perceptually faithful images from severely degraded observations. To address this issue, we propose \textbf{OSD-JSCC}, which couples an SNR-conditioned latent JSCC (L-JSCC) codec with a pretrained one-step diffusion prior. Specifically, a pretrained VAE encoder maps the image to a latent, which the L-JSCC encoder converts into channel symbols. Both the L-JSCC encoder and decoder are conditioned on the SNR through adaptive group normalization (AdaGN). At the receiver, the L-JSCC decoder predicts the mean and variance of the Gaussian state and samples a noisy latent from it, which keeps the channel-induced uncertainty and matches the input expected by the one-step diffusion model. Conditioned on semantic tokens from the L-JSCC decoder and an SNR token, the diffusion U-Net denoises this latent in one step at an SNR-adaptive timestep, and the VAE decoder reconstructs the image. To achieve better performance, we train the JSCC modules end to end and \textcolor{black}{efficiently adapt the diffusion U-Net and VAE decoder} using LoRA. Fig.~\ref{fig:kodim23_awgn_0db} illustrates its perceptual advantage under severe channel noise. Our main contributions are summarized as follows:
\begin{itemize}
 \item We propose OSD-JSCC, an SNR-adaptive latent JSCC framework that integrates a pretrained diffusion prior for high-fidelity image transmission with single-step decoding.

\item We introduce a stochastic diffusion-state interface that predicts a Gaussian latent distribution to model channel-induced uncertainty and extracts image-specific semantic tokens to guide one-step diffusion reconstruction.

\item We develop an SNR-adaptive mechanism that combines feature modulation and diffusion timestep selection to adapt both latent transmission and reconstruction to varying channel conditions.

\item Experiments on Kodak and CLIC2020 demonstrate superior perceptual quality and significantly lower decoding latency than existing diffusion-based JSCC methods at the same bandwidth compression ratio.
\end{itemize}

\section{System Model}
We consider transmitting an RGB image $\bm{x}\in\mathbb{R}^{3\times H\times W}$ over an additive white Gaussian noise (AWGN) channel, where $H$ and $W$ denote its height and width. The encoder maps $\bm{x}$ to a codeword $\bm{s}\in\mathbb{C}^{k}$, where $k$ is the number of complex channel uses. We define the bandwidth compression ratio as $\rho=k/(3HW)$ and normalize the codeword as
\begin{equation}
\bm{s}_c=\sqrt{k}\,\frac{\bm{s}}{\sqrt{\mathbb{E}[\|\bm{s}\|_2^2]}},
\label{eq:power_constraint}
\end{equation}
which satisfies $\mathbb{E}[\|\bm{s}_c\|_2^2]/k=1$. We consider \textcolor{black}{an} AWGN channel, where the received signal can be expressed as
\begin{equation}
\begin{aligned}
\hat{\bm{s}}_c &= \bm{s}_c+\bm{n}_{\mathrm{channel}},\\
\end{aligned}
\label{eq:channel}
\end{equation}
where $\bm{n}_{\mathrm{channel}} \sim \mathcal{CN}(\bm{0},\sigma^2\bm{I})$ is the complex Gaussian noise with zero mean and variance $\sigma^2$. The channel SNR is $\gamma=10\log_{10}(1/\sigma^2)$ in decibels and is assumed available at both ends. The receiver converts the complex observations into real-valued features and reconstructs $\hat{\bm{x}}$. 
\section{Proposed Method}

\subsection{Overview}

Fig.~\ref{fig:osd_jscc_architecture} illustrates the architecture of OSD-JSCC, which integrates a latent JSCC codec with the pretrained VAE and U-Net of SD-Turbo~\cite{rombach2022high,sauer2024adversarial}. At the transmitter, the input image is encoded into a VAE latent and then mapped to channel symbols. At the receiver, the JSCC decoder predicts a stochastic diffusion state and semantic context from the channel observations. The U-Net refines the predicted state in a single step, and the VAE decoder reconstructs the image from the refined latent. To adapt the pretrained generator, we employ low-rank adaptation (LoRA)~\cite{hu2022lora}.

The interface between the JSCC decoder and the pretrained generator is motivated by the diffusion process of SD-Turbo. Given a clean latent $\bm{z}_0$, the noisy state at timestep $t$ is expressed as
\begin{equation}
    \bm{z}_t=\bm{z}_0+\sigma_t\bm{\epsilon},\qquad
    \bm{\epsilon}\sim\mathcal{N}(\bm{0},\bm{I}),
    \label{eq:forward}
\end{equation}
where $\sigma_t$ controls the noise level. The clean latent can then be estimated using a single Euler step as
\begin{equation}
    \hat{\bm{z}}_0=\bm{z}_t-\sigma_t\,
    \epsilon_{\bm{\psi}}\!\left(\bm{z}_t/\sqrt{1+\sigma_t^{2}},\,t,\,\bm{C}\right),
    \label{eq:euler}
\end{equation}
where $\epsilon_{\bm{\psi}}$ denotes the U-Net with LoRA parameters $\bm{\psi}$, and $\bm{C}$ denotes its cross-attention context. To connect the channel observations to the pretrained prior, the JSCC decoder predicts a stochastic state rather than a clean latent. Meanwhile, the timestep is adjusted according to the channel SNR to control the strength of the refinement. \textcolor{black}{These mechanisms are detailed below.}

\subsection{SNR-Conditioned Latent JSCC}

The frozen VAE encoder maps the image to a compact latent
$\bm{z}=\mathcal{E}_{\mathrm{VAE}}(\bm{x})$. A residual adapter
$\mathcal{A}(\bm{z})=\bm{z}+\mathcal{R}_{\bm{\alpha}}(\bm{z})$ adjusts the
pretrained representation for transmission.
The latent JSCC encoder $\mathcal{E}_{\mathrm{L}}$ maps the adapted latent to
channel symbols, and the decoder $\mathcal{D}_{\mathrm{L}}$ extracts a shared
receiver feature $\bm{F}_r$ from the noisy observations. The codec combines
convolutional blocks~\cite{liu2022convnet} with window
attention~\cite{liu2021swin} at the bottleneck to capture local and contextual
information. To adapt the shared codec to varying channel conditions, we embed the SNR
$\gamma$ as $\bm{e}_{\gamma}$ and use adaptive group normalization
(AdaGN)~\cite{wu2018group,dhariwal2021diffusion} in the encoder and decoder:
\begin{equation}
    \operatorname{AdaGN}_{l}(\bm{F},\bm{e}_{\gamma})=
    \operatorname{GN}(\bm{F})\odot\bigl(1+\bm{a}_{l}(\bm{e}_{\gamma})\bigr)
    +\bm{b}_{l}(\bm{e}_{\gamma}),
    \label{eq:adagn}
\end{equation}
where $\bm{a}_l$ and $\bm{b}_l$ project the SNR embedding to the scale and
shift of block $l$. This modulation allows the codec to adjust its features to
the channel quality while sharing network weights across SNRs.

\subsection{Stochastic Diffusion-State Interface}
\label{sec:interface}

The diffusion-state interface maps $\bm{F}_r$ to a stochastic latent and
conditioning tokens for the U-Net. Instead of directly estimating a clean
latent, the decoder predicts a mean $\bm{\mu}$ and a log-variance $\bm{\ell}$
and samples the diffusion state by reparameterization~\cite{kingma2014autoencoding}:
\begin{equation}
    \bm{y}=\bm{\mu}+\exp(\bm{\ell}/2)\odot\bm{\epsilon},\qquad
    \bm{\epsilon}\sim\mathcal{N}(\bm{0},\bm{I}).
    \label{eq:state}
\end{equation}
This construction provides a structured latent with a learned Gaussian
perturbation, motivated by the form of the diffusion state in
\eqref{eq:forward}. The mean and variance are optimized through the
reconstruction losses without an explicit prior or likelihood on them. To guide reconstruction without a text description, we extract semantic tokens
$\bm{C}_{\mathrm{sem}}$ from $\bm{F}_r$ by global pooling and an MLP. We also
project the SNR embedding to a reliability token $\bm{c}_{\gamma}$, forming
the cross-attention context
\begin{equation}
    \bm{C}=[\bm{C}_{\mathrm{sem}};\bm{c}_{\gamma}].
    \label{eq:tokens}
\end{equation}
Both types of tokens are computed at the receiver without additional channel
uses.

\begin{figure*}[t]
\centering
\includegraphics[width=\textwidth]{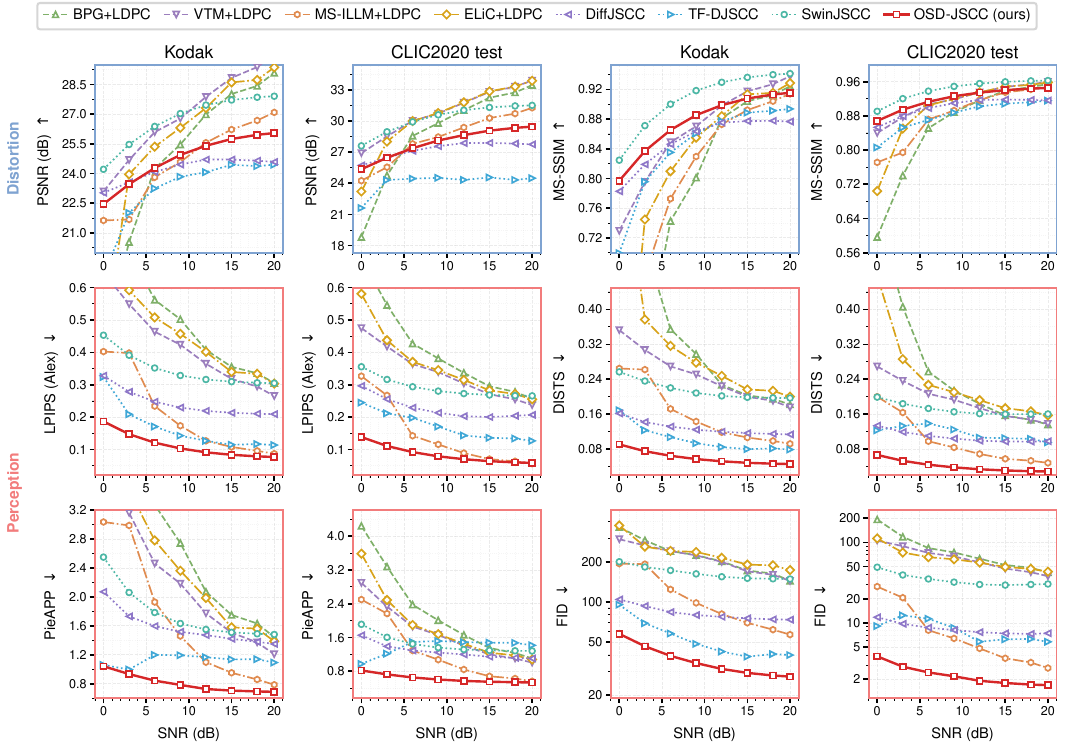}
\caption{Performance comparison of the proposed OSD-JSCC method with baseline methods
on Kodak and CLIC2020 over AWGN at $\rho=1/96$, where SNR ranges from 0 to 20~dB. }
\label{fig:kodak-awgn}
\end{figure*}

The timestep controls the strength of the one-step refinement. We select it
according to the channel SNR:
\begin{equation}
    t(\gamma)=
    \begin{cases}
        249, & \gamma\geq\gamma_{0},\\
        499, & \gamma<\gamma_{0},
    \end{cases}
    \label{eq:routing}
\end{equation}
where $\gamma_0=10$~dB. Both timesteps are drawn from the SD-Turbo sampling
schedule~\cite{sauer2024adversarial}. The lower timestep favors preservation of
the received structure at high SNR, whereas the higher timestep enables
stronger refinement at low SNR. Substituting $\bm{z}_t=\bm{y}$ and
$t=t(\gamma)$ into~\eqref{eq:euler}, we obtain the reconstructed image as
$\hat{\bm{x}}=\mathcal{D}_{\mathrm{VAE}}(\hat{\bm{z}}_0)$ using one U-Net
evaluation.

\subsection{Training Strategy}
Following \cite{aeic}, we adopt a three-stage training strategy to improve reconstruction fidelity and perceptual quality. In Stages~1 and~2, the VAE and U-Net backbones are frozen, except for the U-Net input/output convolutions. We jointly optimize the latent adapter, JSCC codec, SNR embeddings, token projections, and rank-32 U-Net LoRA adapters, totaling 48.2M trainable parameters. During training, the SNR is uniformly sampled from $[0,20]$~dB, with the diffusion timestep selected by~\eqref{eq:routing}. Specifically, Stage~1 establishes reconstruction fidelity using
\begin{equation}
\mathcal{L}_1 = \mathcal{L}_{\mathrm{MSE}} +
\mathcal{L}_{\mathrm{LPIPS}} +
\lambda_{\mathrm{CLIP}}\mathcal{L}_{\mathrm{CLIP}},
\label{eq:stage1_loss}
\end{equation}
where $\lambda_{\mathrm{CLIP}}$ weights the CLIP loss. Stage~2 further improves perceptual quality by introducing the adversarial loss and the DISTS loss.
\begin{equation}
\mathcal{L}_2 = \mathcal{L}_{\mathrm{MSE}} +
\mathcal{L}_{\mathrm{DISTS}} +
\lambda_{\mathrm{adv}}\mathcal{L}_{\mathrm{adv}},
\label{eq:stage2_loss}
\end{equation}
where $\lambda_{\mathrm{adv}}$ controls the adversarial loss. The discriminator employs trainable heads on a frozen DINOv3 backbone~\cite{simeoni2025dinov3}, and training progressively increases the crop size from $512\times512$ to $1024\times1024$. Finally, Stage~3 freezes all preceding modules and fine-tunes the VAE decoder using rank-16 convolutional LoRA adapters with $\mathcal{L}_2$. These adapters are merged after training, adapting the decoder to one-step diffusion outputs without additional inference overhead.

\section{Experiments}

\begin{table}[t]
\centering
\footnotesize
\caption{Running time comparison of different methods on Kodak at 12~dB over AWGN.}
\label{tab:latency}
\newcommand{\latencycell}[3]{#1\,{\scriptsize\textcolor{#2}{(#3$\times$)}}}
\setlength{\tabcolsep}{3pt}
\begin{tabular}{@{}lrrr@{}}
\toprule
Method & Encoding (s) & Decoding (s) & Total (s) \\
\midrule
\multicolumn{4}{@{}l}{\textit{Classical image codec + channel coding}} \\
BPG+LDPC & \latencycell{0.0650}{green!50!black}{2.65} & \latencycell{0.0100}{red!75!black}{6.72} & \latencycell{0.0750}{red!75!black}{1.22} \\
VTM+LDPC & \latencycell{8.0868}{green!50!black}{330.07} & \latencycell{0.0642}{red!75!black}{1.05} & \latencycell{8.1510}{green!50!black}{88.89} \\
\midrule
\multicolumn{4}{@{}l}{\textit{Learned image codec + channel coding}} \\
MS-ILLM+LDPC~\cite{muckley2023improving} & \latencycell{0.0305}{green!50!black}{1.24} & \latencycell{0.0410}{red!75!black}{1.64} & \latencycell{0.0715}{red!75!black}{1.28} \\
\textcolor{black}{ELIC}+LDPC~\cite{he2022elic} & \latencycell{0.0825}{green!50!black}{3.37} & \latencycell{0.0568}{red!75!black}{1.18} & \latencycell{0.1393}{green!50!black}{1.52} \\
\midrule
\multicolumn{4}{@{}l}{\textit{NN-based JSCC}} \\
DeepJSCC~\cite{bourtsoulatze2019deep} & 0.0074 {\scriptsize\textcolor{red!80!black}{(3.31$\times$)}} & 0.0096 {\scriptsize\textcolor{red!80!black}{(7.00$\times$)}} & 0.0172 {\scriptsize\textcolor{red!80!black}{(5.33$\times$)}} \\
SwinJSCC~\cite{yang2024swinjscc} & \latencycell{0.0104}{red!75!black}{2.36} & \latencycell{0.0109}{red!75!black}{6.17} & \latencycell{0.0213}{red!75!black}{4.31} \\
\midrule
\multicolumn{4}{@{}l}{\textit{Generative JSCC}} \\
DiffJSCC~\cite{yang2024diffusion} & \latencycell{\textbf{0.0045}}{red!75!black}{5.44} & \latencycell{3.3994}{green!50!black}{50.59} & \latencycell{3.4038}{green!50!black}{37.12} \\
TF-DJSCC~\cite{shunpu2025enabling} & \latencycell{0.1342}{green!50!black}{5.48} & \latencycell{\underline{0.3829}}{green!50!black}{5.70} & \latencycell{\underline{0.5172}}{green!50!black}{5.64} \\
\rowcolor[HTML]{F3F0FF}
\textbf{OSD-JSCC (ours)} & \underline{0.0245} & \textbf{0.0672} & \textbf{0.0917} \\
\bottomrule
\end{tabular}
\end{table}

\begin{figure*}[!t]
\centering
\includegraphics[width=\textwidth]{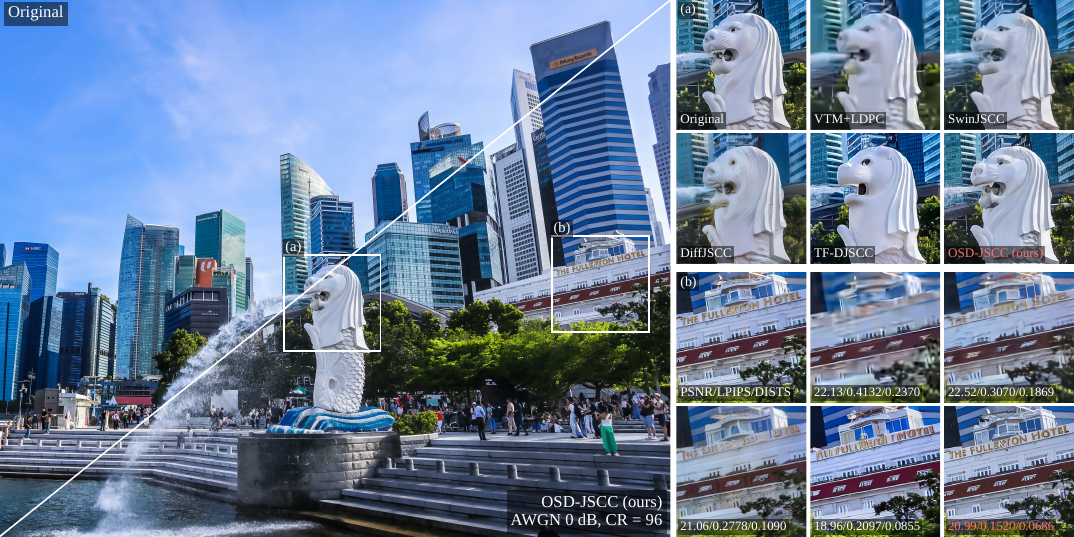}
\caption{Visual comparison on an AWGN channel at
0~dB and compression ratio 96.}
\label{fig:merlion_awgn_0db}
\end{figure*}

\subsection{Setup}

We train the proposed model on 14{,}046 images: 2{,}560 from Flickr2K,
800 from the DIV2K training set~\cite{agustsson2017ntire}, 585 and 41 from the
CLIC2020 professional training and validation sets, respectively, 60 from the
CLIC2021 test set, and the first 10{,}000 from LSDIR~\cite{li2023lsdir}. We
validate on the DIV2K validation set and test on Kodak\footnote{\url{https://r0k.us/graphics/kodak/}} and the
CLIC2020 test set\footnote{\url{https://www.tensorflow.org/datasets/catalog/clic}}. All methods use $\rho=1/96$ over AWGN. We
compare with separation-based schemes combining BPG, VTM,
MS-ILLM~\cite{muckley2023improving}, or ELIC~\cite{he2022elic} with 5G LDPC
codes and QAM, as well as SwinJSCC~\cite{yang2024swinjscc},
DiffJSCC~\cite{yang2024diffusion}, and TF-DJSCC~\cite{shunpu2025enabling}. In Stage~1,
we train for 60k steps using Adam with a batch size of 8, $512\times512$ crops,
and a learning rate of $10^{-4}$ with cosine decay. In Stage~2, we train for
13k steps with $512\times512$ crops and 4k steps with $1024\times1024$ crops at
a learning rate of $5\times10^{-5}$, and average two late checkpoints. Stage~3
uses 3k steps. Training and timing use a single NVIDIA RTX Pro 6000 GPU with 96~GB memory. 

\subsection{Performance Comparison}
In Fig.~\ref{fig:kodak-awgn}, we compare OSD-JSCC with the baselines on Kodak and
CLIC2020 over AWGN. From this figure, we can see that OSD-JSCC achieves the best perceptual quality at all SNRs, with much lower LPIPS, DISTS, and FID scores than the baselines, while maintaining \textcolor{black}{PSNR and MS-SSIM scores comparable to or better than those of} generative JSCC methods, such as DiffJSCC and TF-DJSCC. These results demonstrate the effectiveness of our proposed OSD-JSCC in achieving superior perceptual quality while preserving the fidelity of the reconstructed images.

Table~\ref{tab:latency} reports per-image latency on Kodak at 12~dB. From this table, we can see that the proposed OSD-JSCC decodes in 67~ms, 50.6$\times$ faster than DiffJSCC, whose 50 sampling steps and image captioning take 3.4~s, and 5.7$\times$ faster than TF-DJSCC. The total latency of OSD-JSCC is 92~ms, which is still comparable to that of the separation-based schemes with learned codecs. These results demonstrate the advantages of our proposed 
OSD-JSCC in terms of decoding latency. \textcolor{black}{By using a one-step diffusion model}, OSD-JSCC can achieve a significant reduction in decoding time while maintaining high perceptual quality, making it suitable for real-time image transmission applications.

In Fig.~\ref{fig:merlion_awgn_0db}, we present visual comparisons of the proposed OSD-JSCC and the baselines on the AWGN channel at \textcolor{black}{0~dB}. From this figure, we can see that the proposed OSD-JSCC achieves the best perceptual quality among all the methods, with clearer textures and sharper edges on the Merlion. Moreover, the proposed OSD-JSCC can recover more \textcolor{black}{textual details}, while the other methods fail to recover these details. These results further demonstrate the effectiveness of our proposed OSD-JSCC in achieving superior perceptual quality and preserving the fidelity of the reconstructed images.

\subsection{Ablation Studies}

We conduct ablation studies on Kodak to evaluate the contributions of the proposed modules. All ablations use Kodak and Stage-1 models trained with the same configuration. Table~\ref{tab:module_ablation} compares the full model with two ablations: one without SNR modulation and one without semantic tokens. \textcolor{black}{We report BD-SNR (dB), which measures the average difference in required channel SNR at matched LPIPS or DISTS, using the full model as the reference. From the result we find that removing SNR modulation increases the required SNR by an average of 0.66~dB and 0.58~dB at matched LPIPS and DISTS, respectively. Removing semantic tokens yields BD-SNR values of $+0.25$~dB for LPIPS and $-0.38$~dB for DISTS, indicating a tradeoff between the two perceptual criteria. These results show that SNR modulation improves performance under both criteria, while semantic tokens favor LPIPS at the expense of DISTS.}

\begin{table}[t]
\centering
\small
\caption{Ablation study of the proposed modules on Kodak.}
\label{tab:module_ablation}
\setlength{\tabcolsep}{6pt}
\renewcommand{\arraystretch}{1.12}
\begin{tabular*}{\columnwidth}{@{\extracolsep{\fill}}lcc@{}}
\toprule
 & \multicolumn{2}{c}{\textcolor{black}{BD-SNR (dB)}} \\
\cmidrule(lr){2-3}
Model & \textcolor{black}{LPIPS $\downarrow$} & \textcolor{black}{DISTS $\downarrow$} \\
\midrule
Full & \textbf{0.00} & 0.00 \\
w/o SNR modulation & $+$0.66 & $+$0.58 \\
w/o semantic tokens & $+$0.25 & \textbf{$-$0.38} \\
\bottomrule
\end{tabular*}
\end{table}

\begin{figure}[t]
\centering
\includegraphics[width=\columnwidth]{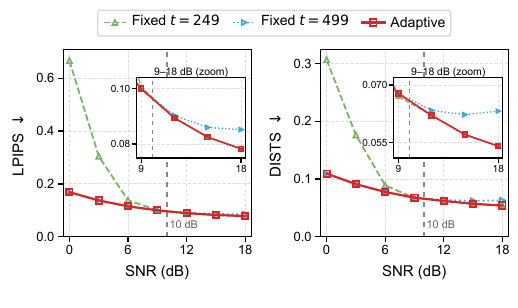}
\caption{Ablation study of the SNR-dependent timestep selection policy on Kodak, while the ablations use a fixed timestep of $t=249$ or $t=499$.}
\label{fig:timestep_policy}
\end{figure}

Fig.~\ref{fig:timestep_policy} shows that routing all SNRs to $t=249$ leaves
results above 10~dB almost unchanged but raises LPIPS at 0~dB from 0.169 to 0.667,
because $\sigma_{249}$ cannot remove the perturbation that the decoder prepared
for $\sigma_{499}$. Routing all SNRs to $t=499$ over-corrects reliable states and worsens DISTS at 18~dB from 0.054 to 0.063. These results demonstrate the effectiveness of the proposed SNR-dependent timestep selection policy in improving the perceptual quality of the reconstructed images.

\section{Conclusion}

In this paper, we have proposed OSD-JSCC, a one-step diffusion-based JSCC codec for wireless image transmission. We have designed an SNR-conditioned latent JSCC codec that performs joint source--channel coding in the VAE latent space, and an SNR-adaptive one-step generation scheme that selects the refinement strength according to the channel SNR. Experimental results have shown that OSD-JSCC achieves superior perceptual quality while maintaining competitive fidelity \textcolor{black}{and significantly reduces decoding latency} compared to existing generative JSCC methods. 

\bibliographystyle{IEEEtran}
\bibliography{IEEEabrv,references}

\end{document}